\documentclass[twocolumn,prb,aps,epsf,amsmath,amssymb,showpacs]{revtex4}
\usepackage{comment}
\usepackage[dvipsnames]{xcolor}
\usepackage{ulem}
\usepackage{yfonts}

\usepackage{graphicx}% Include figure files
\usepackage{dcolumn}% Align table columns on decimal point
\usepackage{bm}% bold math

\newcommand{\nn}{\nonumber}

\def\bea{\begin{equation}\begin{aligned}}
\def\eea{\end{aligned}\end{equation}}

\newcommand{\WMS}[1]{\textcolor{blue}{#1}}
\newcommand{\CS}[1]{\textcolor{magenta}{#1}}

\begin{document}
\title{Onsager Relations between Spin Currents and Charge Currents}
\author{Wayne M. Saslow}
\email{wsaslow@tamu.edu}
\affiliation{ Texas A\&M University, College Station, Texas, 77843, U.S.A. }
\author{Chen Sun}
\email{chensun@hnu.edu.cn}
\affiliation{ School of Physics and Electronics, Hunan University, Changsha 410082, China. }
\begin{abstract}
We consider the macroscopic dynamics of systems with charge and spin currents, using the methods of Onsager's irreversible thermodynamics.  Applied to systems with spin-orbit interaction (SOI), we derive Onsager relations showing that, if electrical disequilibrium leads to spin currents, then magnetic disequilibrium leads to charge currents.  We consider three examples of such SOI.  Two of these predicted charge currents have not previously appeared.  By measuring these charge currents one can infer the corresponding spin currents.

\end{abstract}

\date{\today}

\maketitle

%\tableofcontents
%\textcolor{red}{Chen}
%\textcolor{cyan}{Shanglong}
%\textcolor{blue}{Wayne.

%\section{Introduction}
%\label{s:intro}
\noindent{\bf Introduction.}

Although spin flux (or, effectively, spin current) is essential to the field of spintronics, except for optical techniques that do not apply to metals, spin flux cannot be observed directly.  Typically spin flux is inferred from a measurement like the inverse spin Hall effect, whereby a spin flux leaving a material, with spin polarization in the plane of the surface, produces an SOI-induced charge flux (and thus voltage difference) in the other direction in the plane of the surface. Generally, measurement of charge fluxes is much easier than that of spin fluxes. The main result of this paper is that, for two theoretical models of systems with SOI that predict spin fluxes driven by electrical disequilibrium,
%\WMS{\sout{we find that }}
Onsager reciprocity predicts corresponding charge fluxes driven by magnetic disequilibrium.  It would be of interest to measure such charge fluxes.

In 1971 the effects of SOI was studied in two pioneering works by Dyakonov and Perel, DP1\cite{DyakonovPerel1} and DP2,\cite{DyakonovPerel2} which   for semiconductors considered (dimensionless) charge flux $q_{i}$ and (dimensionless) spin flux $\vec{q}_{i}$.  Later the charge flux was  called the spin Hall effect and the spin flux was called the inverse spin Hall effect; and their theory implied the anomalous Hall effect.\cite{HirschSHE}  DP2,\cite{DyakonovPerel2} submitted a week after DP1,\cite{DyakonovPerel1} gave additional spin flux terms due to previously neglected spin-orbit effects.  These new terms were later expanded on and named ``spin-swapping'' by Lifshits and Dyakonov.\cite{Dyakonov09}  We refer to these works,\cite{DyakonovPerel1,DyakonovPerel2,Dyakonov09} and one by Dyakonov alone,\cite{Dyakonov07} collectively as the {\it DP model.}  DP considered only a non-equilibrium, flux-carrying spin density that they called the {\it accumulation of spin}. Later, in a context without SOI, Valet and Fert \cite{ValetFert93} introduced the term {\it spin accumulation} for this non-equilibrium spin density.  Often spin density $\vec{S}$ and the proportional magnetization density $\vec{M}$ are used interchangeably.

For a ferromagnet with local equilibrium magnetization $\vec{M}$ along $\hat{z}$, versions of the charge flux and spin flux including spin-orbit effects were given by Taniguchi {\it et al} (TGS).\cite{TGS15}  This included the anisotropic magnetoresistances (relative to the magnetization direction $\hat{M}$) associated with gradients of the electrochemical potential $\tilde\mu$ and gradients of only the longitudinal part $\hat{M}\cdot\vec{\mu}$ of the spin electrochemical potential $\vec{\mu}$.\cite{Tserkovnyak05}  Here $\mu$ and $\hat{M}\cdot\vec{\mu}$ are the symmetric and anti-symmetric combinations of the up-spin and down-spin electrochemical potentials.  Ref.~\onlinecite{TGS15} includes the anomalous Hall flux due to both of these gradients, but omits the transverse part of the spin electrochemical potential $\vec{\mu}$.

More recently, Amin {\it et al} (ALSH) gave additional SOI-induced spin flux terms;\cite{ALSH19} they are new, and not analogous to the spin-swapping terms of Dyakonov.\cite{DyakonovPerel2,Dyakonov09} % We find that there are corresponding new charge flux terms, given in \eqref{qiALSH2}.

Because of the role of the SOI in the spin Hall effect and the inverse spin Hall effect -- two powerful spintronics probes -- SOI-related spin currents in ferromagnets have been the subject of ongoing theoretical and experimental interest.  The purpose of the present work is to apply Onsager reciprocity to the three theoretical models with SOI mentioned above, namely, the DP2,\cite{DyakonovPerel2} TGS,\cite{TGS15} and ALSH\cite{ALSH19} models.\cite{note-Onsager}  We show that TGS\cite{TGS15} is consistent with that reciprocity, but that the other two\cite{DyakonovPerel2,ALSH19} must be given charge fluxes that correspond to  the spin flux already in the models.  For the new charge current associated with DP2,\cite{DyakonovPerel2} see eq.~\eqref{delqi}; for the new charge current associated with ALSH,\cite{ALSH19} see eq.~\eqref{qiALSH}.  In a separate section we estimate the size of these effects and propose experiments to measure them.\\ %Sect.\ref{s:exp}.
\indent{}These two new charge currents, together with the proposal to measure them to verify the original spin currents, are the main points of this paper.\\

%We approach the problem using irreversible thermodynamics, with an emphasis on the magnetization $\vec{M}$ and the associated spin current $\vec{J}_{i}$...

%Note the difference between the magnetic systems considered here and $^{3}$He.  $^{3}$He, for which Refs.~\onlinecite{Leggett70,BaymPethick78} present the kinetic theory, is not subject to lattice drag. However, electrons are subject to lattice drag, which leads to diffusive motion.

%However, unlike $^{3}$He, for which Refs.~\onlinecite{Leggett70,BaymPethick78} present the kinetic theory, {\bf electrons in a lattice are subject to lattice drag.  Alternatively, although momentum conservation is replaced by lattice momentum, the latter is conserved only modulo a lattice wavevector.  This leads to diffusion of the lattice momentum.}
%\section{Thermodynamics and Equations of Motion}
%\label{s:thermo&mot}
\noindent{}{\bf Thermodynamics and equations of motion.}

We approach the problem using Onsager's irreversible thermodynamics. We consider charge carriers with    negative charge $-e$ and gyromagnetic ratio $-\gamma$, as normally appropriate for electrons.  They are in an electrical potential $V$, and have chemical potential $\mu$.  The electrochemical potential $\tilde\mu=\mu-eV$.  With $E_{i}=-\partial_{i}V$, we employ the effective electric field
\begin{equation}
E^{*}_{i}=\frac{1}{e}\partial_{i}\tilde\mu=E_{i}+\frac{1}{e}\partial_{i}\mu.
\label{E*}
\end{equation}
%\subsection{Thermodynamics}
%\label{ss:thermo}
{\bf (a) Thermodynamics.} We take the energy density to be a function of the entropy density $s$ (with thermodynamic conjugate the temperature $T$), the carrier density $n$ (with thermodynamic conjugate the electrochemical potential $\tilde\mu=\mu-eV$), and (instead of the magnetization $\vec{M}$) the spin density $\vec{S}$ and the vector number density $\vec{n}$, where
\begin{equation}
\vec{M}=-\gamma\vec{S}=-\frac{\gamma\hbar}{2}\vec{n}.
\label{M-to-n}
\end{equation}
We use units where $B$ is in tesla (T) and both $M$ and $H$ are in A/m: $\vec{B}=\mu_{0}(\vec{H}+\vec{M})$.

Then for the differential of the energy density we take
\begin{equation}
d\varepsilon=Tds+\tilde{\mu}dn+\vec{\mu}\cdot d\vec{n}.
\label{dpsnew}
\end{equation}
Here $\vec{\mu}$ is the spin-space vector chemical potential, as defined in eq.~(16) of %the review by Tserkovnyak et. al.
Ref.~\onlinecite{Tserkovnyak05}, which calls $\vec{\mu}$ the spin accumulation, a term that properly refers to either $\vec{S}$ or $\vec{M}$. \cite{DyakonovPerel1,ValetFert93}

$\vec{\mu}$ is related to the effective magnetic field $\vec{B}^{*}$ (including both the applied field $\vec{B}$ and the internal field, defined so that in equilibrium $\vec{B}^{*}=- \partial\varepsilon/\partial \vec{M} =\vec{0}$).  This is done by writing the (effective) magnetic field energy as
\begin{equation}
-\vec{B}^{*}\cdot d\vec{M}=\vec{\mu}\cdot d\vec{n},
\label{Bdm-to-mudn}
\end{equation}
from which we deduce that
\begin{equation}
\vec{\mu}=\frac{\gamma\hbar}{2}\vec{B}^{*}.
\label{mu-to-B}
\end{equation}
Thus, in equilibrium $\vec{\mu}=\vec{0}$.

Compared with Dyakonov and Perel, the dimensionless charge and spin currents are
\begin{equation}
J^{n}_{i}\equiv q_{i}, \qquad \vec{J}_{i}^{\vec{n}}\equiv\vec{q}_{i}.
\label{DPfluxes}
\end{equation}
The charge and magnetization currents are
\begin{equation}
J^{q}_{i}\equiv-J^{n}_{i}=-q_{i}, \qquad  \vec{J}^{\vec{M}}_{i}\equiv-\frac{\gamma\hbar}{2}\vec{J}^{\vec{n}}_{i}=-\frac{\gamma\hbar}{2}\vec{q}_{i}.
\label{chargspinflux}
\end{equation}

%\subsection{Equations of Motion}
%\label{ss:eqsmot}
{\bf (b) Equations of motion.} With unknown fluxes (we use $J$ rather than $j$, in order to free $j$ to be used as an index) and unknown sources $R$, the equations of motion are taken to be
\begin{eqnarray}
\partial_{t}\varepsilon&+&\partial_{i}J^{\varepsilon}_{i}=0, \label{depsdtnew}\\
\partial_{t}s&+&\partial_{i}J^{s}_{i}=R^{s}\ge0, \label{dsdtnew}\\
\partial_{t}n&+&\partial_{i}J^{n}_{i}=0, \label{dndtnew}\\
\partial_{t}\vec{n}&+&\partial_{i}\vec{J}^{\vec{n}}_{i}=\gamma\vec{n}\times\vec{B}+\vec{R}^{\vec{n}}.\label{dJndtnew}
\end{eqnarray}

We now rewrite $TR^{s}$ in terms of a divergence and of products of thermodynamic fluxes or sources with their respective thermodynamic forces.  To do so we use $\partial_{t}s$, the energy differential, and the other equations of motion.  We find that
\begin{eqnarray}
0\le TR^{s}&=&-\partial_{i}(J^{\epsilon}_{i}-TJ^{s}_{i}-\tilde{\mu}j^{n}_{i}-\vec{\mu}\cdot\vec{J}^{\vec{n}}_{i})\cr
&&-J^{s}_{i}\partial_{i}T-J^{n}_{i}\partial_{i}\tilde{\mu}-\vec{J}^{\vec{n}}_{i}\cdot\partial_{i}\vec{\mu}-\vec{\mu}\cdot\vec{R}^{n}.\qquad
\label{TRsnew}
\end{eqnarray}

%\subsection{Fluxes and Sources}
%\label{ss:fluxsource}
{\bf (c) Fluxes and sources.} Neglecting off-diagonal transport coefficients, the fluxes are
\begin{eqnarray}
J^{s}_{i}&=&-\frac{\kappa}{T}\partial_{i}T, \label{js0}\\
J^{n}_{i}&=&-\frac{\sigma}{e^{2}}\partial_{i}\tilde{\mu}, \label{jn0}\\
\vec{J}^{\vec{n}}_{i}&=&-\frac{\sigma'}{e^{2}}\partial_{i}\vec{\mu}, \label{jvecn0}\\
\vec{R}^{n}&=&-\Gamma\vec{\mu}. \label{Rn0}
\end{eqnarray}
Here $\kappa, \sigma, \sigma'$ all represent dissipative processes (for $s, n, \vec{n}$), and $\Gamma$ represents a decay process.  (For $TR^{s}\ge0$ these are all non-negative.)

We do not employ anisotropy due to the magnetization $\vec{M}$, although each of the dissipative coefficients $(\kappa, \sigma, \sigma', \Gamma)$ can be replaced by a tensor of the form $A_{\parallel}\hat{M}_{\alpha}\hat{M}_{\beta}+A_{\perp}(\delta_{\alpha\beta}-\hat{M}_{\alpha}\hat{M}_{\beta})$.  These represent diffusion and decay processes depending on whether they involve processes along or normal to $\hat{M}$.

%\subsection{Onsager Relations}
%\label{ss:Onsager}
{\bf (d) Onsager relations.} We now consider the implications of the (so far neglected) off-diagonal transport coefficients, such as are well-known in the thermoelectric effect and the corresponding electrothermal effect.  When the off-diagonal transport coefficients correspond to dissipative processes (either diffusion or decay), the cross-terms add; when the off-diagonal transport coefficients correspond to reactive (non-dissipative) processes, the cross-terms cancel (so they do not contribute to the dissipative quantity $R^{s}$).

Thus, when the spin-orbit interaction (SOI) causes $\tilde{\mu}$ and $\vec{\mu}$ to mix, for dissipative processes we have
\begin{equation}
J^{n,D}_{i}\partial_{i}\tilde{\mu}=\vec{J}^{\vec{n},D}_{i}\cdot\partial_{i}\vec{\mu}. \qquad \hbox{(dissipative)}
\label{OnsagerDissnew}
\end{equation}
Likewise, for reactive (non-dissipative) processes we have
\begin{equation}
J^{n,R}_{i}\partial_{i}\tilde{\mu}=-\vec{J}^{\vec{n},R}_{i}\cdot\partial_{i}\vec{\mu}.  \qquad \hbox{(non-dissipative)}
\label{OnsagerReactnew}
\end{equation}
Relation \eqref{OnsagerDissnew} is analogous to what is used to derive the reciprocal relations between heat current driven by electrochemical potential gradients and charge current driven by temperature gradients.  (There is no analog of \eqref{OnsagerReactnew}.) We will apply \eqref{OnsagerDissnew} and \eqref{OnsagerReactnew} to the spin currents of DP,\cite{DyakonovPerel2} TGS,\cite{TGS15} and ALSH \cite{ALSH19} theories.\\

%\section{Weakly spin-polarized system with SOI: Re-writing DP}
%\label{s:rewriteDP}

\noindent{\bf Weakly spin-polarized system with SOI: re-writing DP.}

Before considering ferromagnets, we discuss paramagnets, as considered in the DP model,\cite{DyakonovPerel1,DyakonovPerel2} which we consider to be weakly spin-polarized systems.
Dyakonov and Perel considered the spin flux first and then the charge flux,\cite{DyakonovPerel1,DyakonovPerel2} but we consider them in the opposite order.  %\WMS{\bf \sout{The first two subsections neglect the spin-swapping effects of DP2.} }

In comparing to DP, note that their dimensionless spin polarization $\vec{P}$ is equivalent to the vector number density $\vec{n}$ used above, or $\vec{P}\equiv\vec{n}$.

Only for a paramagnet in equilibrium we define $\chi$ via $\vec{M}=(\chi/\mu_{0})\vec{B}$.  Then, out of equilibrium, we take
\begin{equation}
\vec{B}^{*}=\vec{B}-\frac{\mu_{0}}{\chi}\vec{M}=\vec{B}+\frac{\gamma\hbar}{2}\frac{\mu_{0}}{\chi}\vec{n},
\label{B*P}
\end{equation}
so
\begin{equation}
\vec{\mu}=\frac{\gamma\hbar}{2}\vec{B}+\lambda\vec{n}, \quad \lambda \equiv \frac{\mu_{0}}{\chi}(\frac{\gamma\hbar}{2})^{2}.
\label{vecmuP}
\end{equation}
Also, for a paramagnet
\begin{equation}
\chi=(\frac{\gamma\hbar}{2})^{2}\frac{n}{k_{B}T},
\label{chiP}
\end{equation}
and the charge and spin conductivities are equal: $\sigma=\sigma'$.

%\subsection{Number Flux}
{\bf (a)  Charge flux $q_{i}$.} Following (8.7) of Ref.~\onlinecite{DyakonovReview}, and eliminating gradients of the density by using $E^{*}_{i}$ for $E_{i}$, we have (with $\mu$ the charge mobility)
\begin{equation}
J^{n}_{i}\equiv q_{i}=-\mu n E^{*}_{i}-\beta(\vec{E}^{*}\times\vec{P})_{i}-\delta(\vec{\nabla}\times\vec{P})_{i},
\label{qi}
\end{equation}
where $\beta$ and $\delta$ (which can have either sign) are parameters due to the SOI.  For a nondegenerate semiconductor, $\beta=e\delta/k_{B}T$.\cite{DyakonovPerel1,DyakonovPerel2}  For a paramagnet the conductivity $\sigma=en\mu$.

We now use \eqref{E*}, and \eqref{vecmuP} at fixed $\vec{B}$.  %introduce the new parameter $\lambda = (\mu_{0}/\chi_{m})(\gamma\hbar/2)^{2}.$
Then we may rewrite \eqref{qi} as %(36) of Ref.~\onlinecite{Saslow15} rewrites this as
\begin{equation}
J^{n}_{i}\equiv q_{i}=-\frac{\mu n}{e} \partial_{i}\tilde\mu-\frac{\beta}{e}(\vec{\nabla}\tilde\mu\times\vec{P})_{i}-\frac{\delta}{\lambda}(\vec{\nabla}\times\vec{\mu})_{i}.
\label{qi1}
\end{equation}
The first term is the ordinary conductivity, the second term leads to the anomalous Hall effect, and the third term leads to the inverse spin Hall effect.  Consistent with Onsager's irreversible thermodynamics, the driving terms are gradients of $\tilde\mu$ and $\vec{\mu}$; these gradients are zero in equilibrium.

The second and third terms, proportional to $\partial_{i}\vec{\mu}$, are off-diagonal in the sense of Onsager symmetry, and imply two terms that should appear in the spin current, but proportional to $\partial_{i}\tilde{\mu}$.

%\subsection{Spin Polarization Flux}

{\bf (b) Spin polarization flux $q_{ij}$.} It is not uncommon for ``spin flux'' to refer to spin polarization flux (a density times a velocity), spin flux (an extra factor of $\hbar/2$), or magnetization flux (for electrons, an extra factor of $-\gamma\hbar/2$).

We follow (37) of Ref.~\onlinecite{Saslow15}, which is a rewritten version of (8.8) of Ref.~\onlinecite{DyakonovReview}:
\begin{equation}
(\vec{J}^{\vec{n}}_{i})_{j}\equiv q_{ij}=-\frac{\mu}{e} (\partial_{i}\tilde\mu) P_{j}-\frac{D}{\lambda}\partial_{i}\mu_{j}+\epsilon_{ijk}\frac{\beta n}{e}\partial_{k}\tilde\mu.
\label{qij2}
\end{equation}
For a semiconductor, an Einstein relation gives the diffusion constant
\begin{equation}
D=\frac{\sigma}{e^{2}}\frac{\partial\mu}{\partial n}\approx \frac{\sigma}{e^{2}} \frac{k_{B}T}{n}=\frac{n\mu}{e}k_{B}T.
\label{DEinstein}
\end{equation}

%\subsection{DP1 and Onsager Reciprocity}
{\bf (c) DP1 and Onsager reciprocity.} Eq.~\eqref{TRsnew} contains the terms $-J^{n}_{i}\partial_{i}\tilde{\mu}-\vec{J}^{\vec{n}}_{i}\cdot\partial_{i}\vec{\mu}$, which must be non-negative.  Using \eqref{qi1}, the first term of $-J^{n}_{i}\partial_{i}\tilde{\mu}$ and, using \eqref{qij2}, the second term of $-\vec{J}^{\vec{n}}_{i}\cdot\partial_{i}\vec{\mu}$, are non-negative for $\mu$ non-negative, as expected.  Being quadratic in the respective thermodynamic forces $\partial_{i}\tilde{\mu}$ and $\partial_{i}\vec{\mu}$, Onsager reciprocity does not apply to these terms.

Onsager reciprocity applies to the off-diagonal terms associated with the third terms of \eqref{qi1} and \eqref{qij2}.  These give products of $\partial_{i}\tilde{\mu}$ and $\partial_{i}\vec{\mu}$ with respective coefficients $\delta/\lambda$ and $\beta n/e$, which can be shown to satisfy Onsager reciprocity.  These cross-terms can be of either sign, and thus must not be too large, or $TR^{s}$ can become negative.

The second term of \eqref{qi1}, multiplied by $\partial_{i}\tilde{\mu}$, is identically zero.  The first term of \eqref{qij2}, multiplied by $\partial_{i}\vec{\mu}$, is non-zero, but proportional to the small quantity $\vec{P}$, so we neglect it.  Thus, in the end, the original theory of DP1 satisfies Onsager reciprocity.\cite{DyakonovPerel1}

%\subsection{DP2 Spin-Swapping Spin Flux}
{\bf (d) DP2 spin-swapping spin flux.} A week after giving the first version of the spin flux, DP2 added to the spin flux some additional terms due to the SOI.\cite{DyakonovPerel2}  We have written these as $\Delta q_{ij}$.\cite{Saslow15}  Decades later, Lifshits and Dyakonov called these ``spin-swapping'' terms.\cite{Dyakonov09}

The spin-swapping terms are written implicitly in (3) of Ref.~\onlinecite{Dyakonov09}, which introduces the new spin-swapping parameter $\kappa$.  Ref.~\onlinecite{Saslow15}, in eq.~(38), gives them as
\begin{eqnarray}
(\Delta \vec{J}^{\vec{n}}_{i})_{j}&\equiv& \Delta q_{ij}\cr
&=&-\frac{\kappa\mu}{e}(P_{i}\partial_{j}\tilde\mu-\delta_{ij}\vec{P}\cdot\vec{\nabla}\tilde\mu)
-\frac{\kappa D}{\lambda}(\partial_{j}\mu_{i}-\delta_{ij}\vec{\nabla}\cdot\vec{\mu}).\qquad
\label{Delqij1}
\end{eqnarray}
%where $D$ is the diffusion constant, related to $\mu$ by an Einstein relation.
Ref.~\onlinecite{Saslow15}  uses $\kappa_{s}$ for $\kappa$; these are dimensionless.

The  $\kappa\mu$  %first
terms in the spin-swapping part of the spin flux are diagonal in the matrix of fluxes {\it vs} thermodynamic forces.  However, the  $\kappa D$    terms are off-diagonal, so by Onsager's reciprocity principle they must have corresponding terms in the charge current.  Let us write these   $\kappa\mu$ terms, involving gradients of $\tilde{\mu}$ explicitly:
\begin{equation}
%(\Delta \vec{J}^{\vec{n}}_{i})_j\equiv
\Delta q_{ij}^{\tilde\mu}=-\frac{\kappa\mu}{e}(P_{i}\partial_{j}\tilde\mu-\delta_{ij}\vec{P}\cdot\vec{\nabla}\tilde\mu).
\label{DPqij-od}
\end{equation}
DP2 gives no corresponding charge flux terms, which by Onsager should appear in the theory.

%\subsection{Onsager Gives DP2 A Spin-Swapping Charge Flux}

{\bf (e) Onsager gives DP2 a spin-swapping charge flux.} The intrinsic (and reversible) time signature of $\Delta q_{ij}$ is that of spin times velocity, and thus is even.  On the other hand, irreversible thermodynamics gives \eqref{Delqij1} for $\Delta q_{ij}$, for which the time-signature of each term on the right-hand-side is odd; this is because $\tilde\mu$ is even but $\vec{P}$ and $\vec{\mu}$ are odd.  Thus the intrinsic and irreversible thermodynamics time signatures are opposite, indicating that $\Delta q_{ij}$ is irreversible.  This is consistent with the condition that the dissipation rate $\Delta q_{ij}\partial_{i}\mu_{j}$ be positive and invariant under time-reversal.

By Onsager's reciprocity principle for dissipative terms, \eqref{OnsagerDissnew} then gives
\begin{equation}
%\Delta J^{n}_{i}\equiv
\Delta q^{\vec\mu}_{i}=-\frac{\kappa\mu}{e}(P_{j}\partial_{j}\mu_{i}-P_{i}\partial_{j}\mu_{j}).
\label{delqi}
\end{equation}
This result is implicit, but not commented on, in Ref.~\onlinecite{Saslow15}.\\

%\section{Strongly Spin-polarized system: Spin and Charge Fluxes from TGS and ALSH}
%\label{s:newfluxes}

\noindent{\bf Strongly spin-polarized system: spin and charge fluxes from TGS and ALSH.}

We now present the spin and charge flux from both TGS and ALSH, which are specifically for ferromagnets and, more generally, for strongly spin-polarized systems.

%\subsection{TGS has spin flux and charge flux}
{\bf (a) TGS has spin flux and charge flux.} TGS\cite{TGS15} use the spin flux $Q_{ji}=(\hbar/2)q_{ij}$, with indices opposite of the DP convention that we employ.  From (6) and (7) of TGS\cite{TGS15}, who introduce coefficients $\beta$, $\zeta$, and $\eta$, we write their additional charge flux as $\Delta q_{i}^{\rm T}$ and their additional spin flux as $\Delta q_{ij}^{\rm T}$:
\begin{eqnarray}
\Delta q_{i}^{\rm T}&=& -\frac{\sigma}{e^{2}}\partial_i \{ \tilde\mu +\beta  (\hat{M}\cdot\vec{\mu})\}\nn\\
&&-\frac{\sigma_{\rm AH}}{e^{2}} (\hat{M}\times  \vec\nabla)_{i} \{\tilde\mu +\zeta  ( \hat{M}\cdot\vec{\mu} ) \}\nn\\
&&-\frac{\sigma_{\rm AMR}}{e^{2}} \hat{M}_{i} (\hat{M}\cdot\vec{\nabla}) \{  \tilde\mu  +\eta (\hat{M}\cdot\vec{\mu}) \},
\label{qiTGS}\\
\Delta q_{ij}^{\rm T}
&=& -\frac{\sigma }{e^{2}}\hat{M}_j \partial_{i} \{(\hat{M}\cdot\vec{\mu}) +\beta  \tilde\mu \} \nn\\
&&-\frac{\sigma_{\rm AH}}{e^{2}} \hat{M}_{j}(\hat{M}\times  \vec{\nabla})_{i}  \{ ( \hat{M}\cdot\vec{\mu} ) +\zeta  \tilde\mu \}_{i} \nn\\
&&-\frac{\sigma_{\rm AMR}}{e^{2}} \hat{M}_{i} \hat{M}_{j} (\hat{M}\cdot\vec{\nabla}) \{(\hat{M}\cdot\vec{\mu}) +\eta \tilde\mu \}.\qquad
\label{qijTGS}
\end{eqnarray}
The off-diagonal terms in $\Delta q_{i}^{\rm T}$ and $\Delta q_{ij}^{\rm T}$ for which Onsager relations apply are those proportional to $\beta$, $\zeta$, and $\eta$.   [We neglected these terms in \eqref{jn0} and \eqref{jvecn0}.]  %Sect.~\ref{ss:fluxsource}.)
Indeed they satisfy the appropriate Onsager relations.

Because TGS omits transverse spin diffusion, this $\Delta q_{ij}\equiv \Delta (\vec{q}_{i})_{j}$ includes no terms normal to $\hat{M}_{j}$. Note that the third set of terms in \eqref{qiTGS} and in \eqref{qijTGS} can be thought of as due to ``tensorization'' of the gradients of its first set of terms.

%\subsection{ALSH has only spin flux}

{\bf (b) ALSH has only spin flux.} ALSH find a new spin flux term, which in its general form is given in their footnote [16].\cite{ALSH19}  They express their results using an orthonormal triad of unit vectors $\hat{n}^{i}$, where $\hat{n}^{i}_{j}=\hat{n}^{i}\cdot\hat{n}^{j}=\delta_{ij}$.  Their symbol $\sigma$, which is associated with spin-scattering, is {\it not} the conventional conductivity; to avoid confusion we employ $\tilde\sigma$.

On rewriting their spin flux in tensor form we find
\begin{eqnarray}
\Delta (\vec{q}^{\rm A}_{i})_{j} \equiv \Delta q^{\rm A}_{ij}
&=&\frac{\tilde\sigma_{\parallel}-\tilde\sigma_{\perp}}{e^{2}}\hat{M}_{j}\hat{M}_{k}\varepsilon_{ikl}\partial_{l}\tilde\mu
+\frac{\tilde\sigma_{\perp}}{e}\varepsilon_{ijl}\partial_{l}\tilde\mu.\qquad
\label{qijALSH}
\end{eqnarray}
This $\Delta q^{\rm A}_{ij}$ is driven by $\partial_{i}\Delta \tilde\mu$, which means that the Onsager coefficients are off-diagonal.  Therefore, there must be corresponding $\Delta q^{\rm A}_{i}$ terms driven by $\partial_{i}\vec\mu$. The first term above has the symmetry of the $\zeta$ term in \eqref{qijTGS}. The second term corresponds to the last (spin Hall effect) term in \eqref{qij2}.

%\subsection{Onsager reciprocity and ALSH spin flux implies ALSH charge flux}

{\bf (c) Onsager reciprocity and ALSH spin flux implies ALSH charge flux.} The time-signature of $q^{\rm A}_{ij}$ is even, and each of the new terms in \eqref{qijALSH} is even under time-reversal.  Therefore the new terms are non-dissipative, and do not contribute to the rate of entropy production. Thus \eqref{OnsagerReactnew} applies.  \\

From \eqref{OnsagerReactnew}  we deduce that
\begin{equation}
\Delta q^{\rm A}_{i}=-\frac{1}{e^{2}}[(\tilde\sigma_{\parallel}-\tilde\sigma_{\perp})\hat{M}_{l}\hat{M}_{k}\varepsilon_{ijk}
+\tilde\sigma_{\perp}\varepsilon_{ijl}]\partial_{j}\mu_{l}.
\label{qiALSH}
\end{equation}
For $\hat{M}=\hat{z}$, so $\hat{M}_{k}=\delta_{kz}$, this gives
\begin{equation}
\Delta q_i^{\rm A}=-\frac{1}{e^2} [(\tilde\sigma_\parallel-\tilde\sigma_\perp) \varepsilon_{ijz}\partial_j\mu_z+ \tilde\sigma_\perp \varepsilon_{ijl}\partial_j\mu_l ].
\label{qiALSH2}
\end{equation}
The first term and the second term with $l=z$ together give a contribution $q_{i}=-(1/e^{2})\tilde\sigma_{\parallel}\varepsilon_{ijz}\partial_{j}\mu_{z}$, and the second term with $l=(x,y)$ gives a contribution $q_{i}=-(1/e^{2})\tilde\sigma_{\perp}\varepsilon_{ijl}\partial_{j}\mu_{l}$.  The sum is thus
\begin{equation}
\Delta q_i^{\rm A}=-\frac{\tilde\sigma_{\parallel}}{e^2}\varepsilon_{ijz}\partial_j\mu_z
-\frac{\tilde\sigma_{\perp}}{e^2}[\varepsilon_{ijx}\partial_j\mu_x + \varepsilon_{ijy}\partial_j\mu_y].
\label{qiALSH3}
\end{equation}

%\section{Experimental Consequences}
%\label{s:exp}
\noindent{\bf Experimental consequences for ALSH and DP2.}

\begin{figure}[!htb]
  \scalebox{0.5}[0.5]{\includegraphics{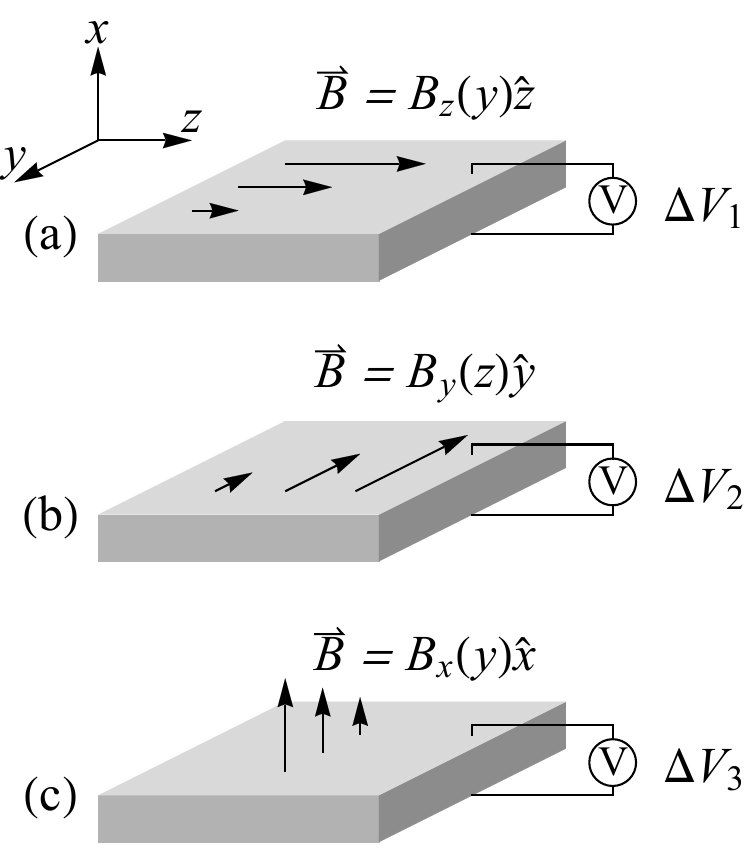}}
%\hspace{-2mm}\vspace{-4mm}%!!
\caption{%(a) Undirected graph of a 4-loop.
Experimental geometries for measuring voltages corresponding to the new charge currents that will verify the existence of the spin currents predicted by: (a) and (b) the ALSH model, and (c) the DP model.} \label{fig:exp-setup}
\end{figure}

{\bf (a) ALSH.}
%\subsection{ALSH}
This charge current can be measured. Consider a thin sample with dimensions $l_x, l_y, l_z$, where $l_x\ll l_y, l_z$ (as shown in Fig.~\ref{fig:exp-setup}).  Recall that, by \eqref{mu-to-B} $d\vec{\mu}=(\gamma\hbar/2)d\vec{B}^{*}\approx (\gamma\hbar/2) d\vec{B}$.  According to \eqref{qiALSH3}, if we apply a field $\vec{B}=B_z(y) \hat z$, then a charge current along $x$, of $\Delta q_{x,1}^{\rm A}=-(\tilde\sigma_{\parallel}/e^2) \partial_y\mu_z$, will be generated.

Let us assume isotropic charge conductivity $\sigma$.  If there are no leads along $\hat{x}$, then the equilibrium $\Delta V_1$ along $x$ is determined by the condition that the charge current $\Delta q_{x,1}^{\rm A}$ is canceled by the conventional dissipative charge current $\sigma_x \Delta V_1/l_x$.  Thus
\begin{align}\label{DeltaV1}
\Delta V_1  = -\frac{ l_x\tilde\sigma_{\parallel}}{ e \sigma}\partial_y\mu_z=-\frac{\tilde{\sigma}_{\parallel}}{\sigma}\frac{\gamma\hbar}{2e}l_{x}\partial_{y}B_{z},
\end{align}
where the negative sign means that the electric potential at the $x=l_x$ surface is higher than that at the  $x=0$ surface.  See Fig.~\ref{fig:exp-setup}(a).

Similarly, if a magnetic field of the form $\vec{B}=B_{y}(z) \hat y$ is applied, then a charge current along $x$, of $\Delta q_{x,2}^{\rm A}=(\tilde\sigma_{\perp}/e^2) \partial_z\mu_y$, will be generated, and there should be a voltage $\Delta V_2$ along $x$:
\begin{align}\label{DeltaV2}
\Delta V_2  = \frac{ l_x\tilde\sigma_{\perp}}{ e \sigma}\partial_z\mu_y=\frac{\tilde{\sigma}_{\perp}}{\sigma}\frac{\gamma\hbar}{2e}l_{x}\partial_{z}B_{y}.
\end{align}
Therefore, measurements of finite $\Delta V_1 $ and $\Delta V_2 $ will test the present theory and determine the values of $\tilde\sigma_{\parallel}$ and $\tilde\sigma_{\perp}$.  See Fig.~\ref{fig:exp-setup}(b).

We now estimate the magnitude of the magnetic field gradient needed to produce a measurable voltage $\Delta V$. From Table I of Ref.~\onlinecite{ALSH19}, values of $\tilde\sigma_{\parallel}$ and $\tilde\sigma_{\perp}$ for Fe, Ni, and Co range from $10^2$ $\Omega^{-1}$ cm$^{-1}$ to $2\times 10^3$ $\Omega^{-1}$ cm$^{-1}$.  We take $\sigma=1.0\times 10^5$ $\Omega^{-1}$ cm$^{-1}$, which is not far from the values for Fe, Ni, and Co, and $l_x=1$ mm.  Then to produce a voltage $|\Delta V|=100$ $\mu$V, we estimate the field gradient $\partial_{y}B_{z}$ or $\partial_{z}B_{y}$ to be from $0.086$ T/$\mu $m to $1.7$ T/$\mu $m.

%{\color{magenta} \bf It is 86 to 1700. I've rewritten it to avoid ambiguities.]}

%\subsection{DP}
%\label{s:DP}

{\bf (b) DP2.} We now discuss possible experimental observation of the spin-swapping charge currents \eqref{delqi} from DP. Again consider a thin sample with dimensions $l_x, l_y, l_z$, where $l_x\ll l_y, l_z$. From \eqref{delqi}, we have
\begin{equation}
\Delta q_{x}=-\frac{\kappa\mu}{e}(P_{y}\partial_{y}\mu_{x}+P_{z}\partial_{z}\mu_{x}-P_{x}\partial_{y}\mu_{y}-P_{x}\partial_{z}\mu_{z}).
\label{}
\end{equation}
If $\vec{P}=P \hat y$ and we apply a field $\vec{B}=B_x(y) \hat x$, then a charge current $\Delta q_{x}=-\frac{\kappa\mu}{e} P \partial_{y}\mu_{x} $ will be generated, and analysis similar to that for ALSH predicts a voltage:
\begin{align}\label{DeltaV3}
\Delta V_3  = -\frac{ l_x\kappa\mu P}{   \sigma }\partial_y\mu_x=-\frac{\kappa\mu P}{\sigma}\frac{\gamma\hbar}{2 }l_{x}\partial_{y}B_x.
\end{align}
See Fig.~\ref{fig:exp-setup}(c).

We take $P=0.01 n$, where $n$ is the charge carrier density and note that $\mu/\sigma=1/(ne)$. %mobility $\mu=10^4$ cm$^2$/(V$\cdot$s)
Then $\Delta V_3 =-0.01 \kappa  \gamma\hbar/(2e )  l_{x}\partial_{y}B_x $. Lifshits and Dyakonov \cite{Dyakonov09} estimate $\kappa$ to be $0.3$ for InSb (large SOI) and $0.003$ for GaAs (small SOI).  Thus, to produce a voltage $|\Delta V_3|=100$ $\mu$V for a $l_x=1$ mm sample, we estimate $\partial_{y}B_x=0.57$ T/$\mu $m for $\kappa=0.3$  and $\partial_{y}B_x=57$ T/$\mu $m for $\kappa=0.003$.\\

%{\color{blue}  Fig.~\ref{fig:exp-setup} shows experimental setups for measuring the voltages $\Delta V_1 $, $\Delta V_2 $ and $\Delta V_3 $ discussed above. Experimental observation of these voltages will confirm the existence of the charge currents and thus their corresponding spin currents in the ALSH and DP2 models. }

%\section{Summary and Conclusions}
%\label{s:summary}
\noindent{}{\bf Summary and conclusions.}\\
We have developed the irreversible thermodynamics of magnets supporting spin currents and charge currents.  We specifically showed that Onsager relations imply that, when there are spin currents driven by voltage gradients, there are also charge currents driven by field gradients, which may enable measurement of effects predicted by Dyakonov and Perel\cite{DyakonovPerel2} and by Amin {\it et al}.\cite{ALSH19} Observation of these charge currents would corroborate the spin currents predicted by Ref.~\onlinecite{DyakonovPerel2} and Ref.~\onlinecite{ALSH19}.\\

%\section{Acknowledgements}
%\label{s:acknow}
\noindent{}{\bf Acknowledgements.}\\
C.S. was supported by National Natural Science Foundation of China under No. 12105094 and by the Fundamental Research Funds for the Central Universities from China.

{}


\begin{thebibliography}{11}
\bibitem{note-Onsager}
This reciprocity holds for any model where the spin current arises primarily from a single energy band, in which case the spin chemical potential is associated with that band.
\end{thebibliography}
\end{document}